\documentstyle[12pt]{article}
\begin{document}
\renewcommand{\thesection}{\Roman{section}}
\baselineskip 20pt
{\hfill PUTP-96-28}
\vskip 5cm
\begin{center}
{\large\bf What about the contribution from reggeons in hard process of strong
interactions?\footnote{The project is supported in part by National Science
Foundation of China,    Doctoral Program Foundation of Institution of Higher
Education of China.}}
\end{center}
\vskip 7mm
\centerline{Hong-An Peng$^{a,b}$,~Shi-Hua Zhao$^a$,~and~ 
Cong-Feng Qiao$^{a,b}$}
\centerline{\small\it $a$.Peking University, 
            Beijing 100871, People's Republic of China}
%\centerline{ Kuang-Ta Chao}
\centerline{\small\it $b$.CCAST
(World Laboratory), Beijing 100080, P.R. China}
%\centerline{\small\it and Department of Physics, Peking University, 
%            Beijing 100871, P.R. China}
\begin{abstract}
%\begin{center}
%\begin{minipage}{130mm}
%\vskip0.6in
%\begin{center}{\bf Abstract}\end{center}

  {We discuss in this paper the possible contributions
  from $I\!\! R_\rho$ ($\rho$-reggeon associated with $\rho$ meson)
in DIS process on protons. Using results
from phenomenological analysis of high energy $\pi$-N charge-exchanged 
scattering process, we get the expression of effective propagator of
$I\!\!R_\rho$ and the  coupling constant $\beta_\rho$ between $I\!\!R_\rho$
and light quarks. As a simple and concrete 
example, we use them to evaluate 
the contributions of $I\!\!R_\rho$ to $F_2(x,Q^2)$ 
from charge-exchanged photoproduction quark-pair process,
$\gamma^*+p\rightarrow n + q_u + \bar{q}_d (\rightarrow n + X)$,
in HERA kinematical regions.
For a comparison, we also evaluate contributions both from pomeron and 
partons in the same 
process. The ratio of contributions in $F_2(x,Q^2)$ from
$I\!\!R_\rho$ versus $I\!\!P$ is larger  than  $1\%$,
 therefore, the former  may be tested experimentaly.
  }
%\vskip 1cm
%PACS number(s):$12.38.Bx$, $~13.38.Dg$, $~14.40.Gx$
%\end{minipage}
%\end{center}
\end{abstract}

%\vfill\eject\pagestyle{plain}\setcounter{page}{1}
\begin{center}
\section{Introduction}
\end{center}
\par
It is well known in high energy strong interactions that the regge pole
theory is the sole one which has both correctly explained and successfully
predicted a lot of soft physics phenomena \cite{1a}. The entities playing 
most important roles in this phenomenological theory are pomeron, 
$I\!\!P$, ( a regge trajectroy with vacuum quantum number) and reggeons,
$I\!\!R$, (regge trajectory with quantum numbers  
of various hadrons). In recent years
since the observations of hard diffractive scattering events in high energy
P-P collisions were reported \cite{2a}, physicists got excited and tried 
with great interesting to explain these phenomena from QCD
and to study the partonic structure of pomeron. Much progress both from
experimental and theoretical research has been achieved and several models of
pomeron have been proposed \cite{3}. But so far as we know nothing 
has been done on this way for the reggeons. Perhaps the following things
could account for the occurrence:
\begin{enumerate}
\item From the  Chew-Frautschi plot of regge trajectories as shown in 
Fig. 1, it can be easily understood that in genuine elastic  scattering
process, contributions of reggeons are always overwhelmed by that of 
pomeron.
\item Several reggeons with different quantum numbers could often be exchanged
simultaneously in the $t$ (or $u$) channel of a single scattering
amplitude, thus the phenomenological analysis for reggeons is rather
difficult.
\item It seems mysterious  about {\it duality} in strong interactions
which is intimately related with reggeons but irrelevant to pomeron, so 
perhaps the dynamical 
structure of reggeons is more complicated than that of pomeron. 
\end{enumerate}

So the
study of partonic structure of reggeons and its QCD explaination is
indeed a
serious and difficult subject. As a primary and simple test of this problem,
we first start from phenomenology of $\pi$-N scattering and try to derive from
them the effective propagator of one single reggeon. For this
purpose, the ideal 
testing place is the high energy charge-exchanged $\pi$-N scattering process,
\begin{eqnarray}
\pi^- + P \rightarrow \pi^0 + n,
\end{eqnarray}
in which only the charged reggeon $I\!\!R_\rho$
 with quantum number $J^{PC}=1^{--}$
(the same as that of $\rho^+$ meson) could be exchanged in its t-channel.
Then, we want to discuss and evaluate the contributions of reggeon 
$I\!\!R_\rho$
in some high energy hard process. The ideal candidate for this purpose
is the charge-exchanged  photoproduction quark-pair
process
\begin{eqnarray}
\label{2}
\gamma^* + P \rightarrow n + q_u + \bar{q}_d (\rightarrow n + X),
\end{eqnarray}
which may be implemented on HERA of DESY. 
 It is special relevant to our aim
that
the  ZEUS Collab. of DESY has installed recently the forward proton
spectrometer (FPS) and the  forward neutron calorimeter (FNCAL) \cite{4a},
which perhaps can be used to detect the  neutron in final state like
Eq.(\ref{2})
process. Thus
with the propagator and coupling constant obtained, we evaluate
the contributions of $I\!\!R_\rho$
in $F_2(x,Q^2)$ for Eq.(\ref{2}). In order for 
comparison we also
evaluate the contribution from $I\!\!P$,
\begin{eqnarray}
\gamma^* + P \rightarrow P +q +\bar{q} (\rightarrow n + X),
\end{eqnarray}
and from $\gamma^* + P \rightarrow X$ mediated by hard subprocess
\begin{eqnarray}
\gamma^* + q \rightarrow q,~~ q+g,~~ \gamma^*+g\rightarrow q +\bar{q},
\end{eqnarray}
respectively. 

Their results are shown  in final section with some comments,
the ratio of contributions in
$F_2(x,Q^2)$ from $I\!\!R_\rho$ versus $I\!\!P$ 
is larger than  $1\%$, therefore the former may be tested  experimentaly.

\begin{center}
\section{Charge-exchanged  $\pi$-N scattering and $\rho$-reggeon
phenomenology}
\end{center}
\par
~~~
\par

It is well known that the transition amplitude of $\pi$-N scattering is
(Fig. 2):
\begin{eqnarray}
A=\bar{u}(p_4) M u(p_2),
\end{eqnarray}
the matrix $M$ is usually written as 
\begin{eqnarray}
M=A(s,t) + \frac{1}{2} (\not\!\!p_2 +\not\!\!p_4) B(s,t)
\end{eqnarray}
A, B are scalar functions of Mandelstam variales $s$ and $t$, $s+t+u=
2(m_{\pi}^2+m_{N}^2)$. When $s\gg m_N^2$,  scalar 
functions A, B would be expressed as the sum of Regge-pole terms which 
could be exchanged in $t$-channel. In order to remove the kinematical
singularities in reggeized expressions, we introduce $A'$ instead of
$A$,
\begin{eqnarray}
A'=A+\frac{E_L-\frac{t}{4 m_N}}{1-\frac{t}{4 m_N^2}}B\equiv
A-m_N z_t \sqrt{\frac{4 m_\pi^2 -t}{4 m_N^2 -t}} B
\end{eqnarray}
where $E_L=\frac{1}{4 m_N}(s+t-u)\approx \frac{S}{2 m_N}$ is the  energy of
$\pi$-N in Lab. system,
$$Z_t\equiv cos\theta_t= \frac{s-u}{\sqrt {(4 m_\pi^2-t)
(4 m_N^2-t)}}=\frac{4 m_N (E_L-\frac{t}{4 m_N})}{\sqrt{(4 m_\pi^2 -t)
(4 m_N^2-t)}}.$$

For charge-exchanged process Eq.(1), only charged $\rho$-regge trajectory
with quantum numbers $J^{PC}=1^{--}$ (the same as that of $\rho$-meson) 
could be exchanged in the $t$-channel. The scalar functions $A^\prime$, $B$ 
expressed in terms of parametric regge formalism have
been given by Rarita et al. \cite{5a}:
\begin{eqnarray}
A'(s,t)=\sqrt{2} C_0 [ e^{C_1 t} (1 + C_2) - C_2] (\alpha_\rho(t)+1)
(\frac{s}{2 m_N^2})^{\alpha_\rho(t)} (\frac{1-e^{-i\pi\alpha_\rho(t)}}
{sin\pi \alpha_\rho(t)}),\\
B(s,t)=\sqrt{2} D_0 e^{D_1 t} \alpha_\rho(t)(\alpha_\rho(t)+1)
(\frac{s}{2 m_N^2})^{\alpha_\rho(t)-1} (\frac{1-e^{-i\pi\alpha_\rho(t)}}
{sin\pi\alpha_\rho(t)}),
\end{eqnarray}
where $\alpha_\rho(t)$ is the trajectory function of $I\!\!R_\rho$ 
 and $\xi_\rho\equiv(\frac{1-e^{-i\pi\alpha_\rho}}
{sin\pi\alpha_\rho(t)})$ is the signature of $\rho$-reggeon 
 from phenomenological analysis, the
 set of parameters $C_i$, $D_i$ 
 were given in ref.\cite{5a}, here we list them in Table 1.
\vskip +4.0mm
\begin{centerline} {Table 1. The magnitudes of parameters $C_i$ and
$D_i$ }\end{centerline}
\vskip 0.5cm 
\begin{center}
\begin{tabular}{|c|c|c|c|c|c|}\hline
 &$C_0(mbGeV)$&$C_1(GeV^{-2})$&$C_2$&$D_0(mb)$&$D_1(GeV^{-2})$\\ \hline
 1&1.47&0.20&15.2&26.3&0.34\\ \hline
 2&1.51&2.39&1.48&29.4&0.14 \\ \hline
 3&1.57&2.02&1.52&29.1&0.11\\ \hline
\end{tabular}
\end{center}

The differencial and total cross section for Eq.(1) may be obtained
from above formulas. 
\begin{eqnarray}
\frac{d\sigma}{dt}(s,t)=\frac{1}{p s}(\frac{m_N}{4 k})^2
[(1-\frac{t}{4 m_N^2}){}|A'|^2 - \frac{t}{4 m_N^2}(\frac{4 m_N^2 p^2 +s t}
{ 4 m_N^2 -t}) |B|^2],
\end{eqnarray}
\begin{eqnarray}
\sigma^{\pi^-p\rightarrow \pi^0 n}(s)=\frac{1}{p} Im A'(s,0),
\end{eqnarray}
where 
\begin{eqnarray} 
p\equiv p^\pi_{Lab.}\stackrel{s\gg m_N^2}{\Large \sim}\frac{s}{2 m_N},~~
k\equiv p_{c.m.}^{\pi}=\frac{s + m_\pi^2 -m_N^2}{2\sqrt{s}}
\stackrel{s\gg m_N^2}{\Large \sim} \frac{\sqrt{s}}{2}.
\end{eqnarray}
With those parameters shown in Table 1, ref. \cite{6a} has shown 
that the 
results from Eqs.(8), (9) and (10)  
fitted with relevant data very well over a large
energy range.
In order to extract the coupling constants between $I\!\!R_\rho$ and hadrons, we should
rewrite the transition amplitude in terms of $s$-channel helicity amplitudes.
Since from above formulas and coefficients $C_i$, $D_i$, everything
is known and ready, thus this is just a matter of converting manipulations.
The $s$-channel helicity amplitudes $A_{Hs}$ and the differential
cross sections for large s are [1]
\begin{eqnarray}
A_{Hs}^{}(s,t) \stackrel{s\gg m_{N}^{2}}{\Large \sim} (\frac{-t}
{2m_N^2})^{\frac{1}{2} m} (\frac{1-e^{-i\pi\alpha_\rho(t)}}
{sin\pi \alpha_\rho(t)}) (\frac{s}{2m_N^2})^{\alpha_\rho(t)}\gamma_\pi(t)
\gamma_{\mu_2\mu_4}(t),
\end{eqnarray}
where
%\begin{eqnarray}
$$m\equiv|\mu_2-\mu_4|$$
%\end{eqnarray}
\begin{eqnarray}
\frac{d\sigma}{dt}(s,t)=
\frac{1}{64 \pi s k^2}\frac{1}{2}\sum|A_{Hs}^{}(s,t)|^2.
\end{eqnarray}
$\mu_2$, $\mu_4=\pm\frac{1}{2}$ are helicities of the initial proton
and final neutron, and $\gamma_\pi(t)$, $\gamma_{\mu_2\mu_4}(t)$
are vertex functions of $I\!\!R_\rho\pi\pi$ and $I\!\!R_\rho NN$,
 respectively. We are interested
in $|t|\approx 0$, so only the m= 0 case in Eq.(13) is considered.

From Eqs.(7)-(11), we get the expression for the 
residue function $\gamma_\pi(t)$
$\gamma_N(t)$ of $I\!\!R_\rho$ in the Eq.(1) process, 
\begin{equation}
\gamma_{\pi}(t)\gamma_N(t)=\frac{\sqrt{2}}{m_N}C_0[(1+C_2)e^{C_1t}-C_2]
(\alpha_{\rho}(t)+1),
\end{equation}
where $\gamma_N(t)\equiv\gamma_{
\frac{1}{2}\frac{1}{2}}(t)=\gamma_{\frac{-1}{2}\frac{-1}{2}}(t)$.

Now let us give a brief discussion about the interaction between reggeon
and partons. Since all regeons are colour singlet and have definite
$I$, $S$, $C$, $B$, etc quantum numbers, they could not couple with 
gluon. As for the coupling between reggeon and quarks, neither model
build-up nor experimental observation yet has been done about partonic
structure of reggeon, so we can only get access to this problem from strong
interaction phenomenologies. In related to this case, we shall make following
ansatzs:
\begin{enumerate}
\item The couplings between reggeon and hadrons, like that in pomeron case,
could be described as a reggeon incoherently coupling with each quark in the
hadron. These coupling constants have $SU_2$ (isotopic spin) symmetry.
\item Regge pole residues could be factorized into a contribution to each 
vertex. For $\pi^-p\rightarrow \pi^0 n$ process, residue function $\gamma^{}(t)$ could be written as 
$$\gamma^{I\!\!R_\rho}(t) = \gamma^{I\!\!R_\rho}_{\pi\pi}\gamma^
{I\!\!R_\rho}_{PN}(t),$$
where $\gamma^{I\!\!R_\rho}_{\pi\pi}$, $\gamma_{NN}^{I\!\!R_\rho}$ are 
coupling vertex functions of $I\!\!R_\rho$ with $\pi$, $N$ respectively. 
\item The duality effect manifested in hadron-hadron interactions
should be embodied by Zweig allowed diagrams, this would prescribe the possible
linkage ways between $I\!\!R$ and quark lines.
\end{enumerate}
On the basis of these assumptions, we can deduce following relations:
\begin{eqnarray}
\label{16}
\gamma_{\pi\pi}^{I\!\!R_\rho}(t)=\beta_\rho(t) F_\pi(t),~~\gamma_{NN}^{I\!\!
R_\rho}(t)=2 \beta_\rho(t) F_N(t),
\end{eqnarray}
where $\beta_\rho(t)$ is the coupling function between $I\!\!R_\rho$
 and light ($u$,
$d$) quarks, $F_\pi(t)$ and $F_N(t)$ are normalized form factors for a
quark in $\pi$-meson and nucleon respectively. Factor 2 in the second part of
Eq.(\ref{16}) comes
from the proton ($uud$) which has two up-quarks. Put Eq.(\ref{16}) into Eq.(12), we get
\begin{eqnarray}
\beta_\rho^2 (t) = \frac{\sqrt{2}C_0 [(1+c_2)e^{c_1 t} -c_2](\alpha_\rho(t) +1)}
{2 m_N F_\pi(t) F_N(t)},~~(t\le0)
\end{eqnarray}
and from Eq.(11)
\begin{eqnarray}
\sigma_T^{\pi^- p\rightarrow \pi^0 n} (s) = 2 \beta_\rho^2 (\frac{s}{2 m_N^2})^
{\alpha_\rho(0)-1}.
\end{eqnarray}

From Eq.(17) and Table 1, we get $\beta_{\rho}$=2.038, 2.066 and 2.107 
$GeV^{-1}$ 
which correspond to the three sets ot parameters in Table 1. We shall
take their average value, i.e. $\beta_{\rho}\equiv\beta_
{\rho}(t)|_{t=0}$=2.070 Ge$V^{-1}$ in our following calculations.
The $\rho$-trajectrory, $\alpha_\rho(t)$, could be estimated from Chew-
Fracitschi plot,
$$\alpha_\rho(t)=\alpha_\rho(0) + \alpha_\rho^\prime t,~~\alpha_\rho(0)
\approx0.5,~~\alpha_\rho^\prime\approx0.9 GeV^{-2}.$$
Thus the effective  $\rho$-reggeon propagator radiated from nucleon
is
\begin{eqnarray}
D_{I\!\!R_\rho}(t)= 2 \beta_\rho^2 F_N(t) (\frac{s}{2 m_N^2})^{\frac{1}{2}(\alpha_
\rho(t) -1)}(\frac{1-e^{-i\pi\alpha_\rho(t)}}{sin\pi\alpha_\rho(t)}).
\end{eqnarray}
For comparing and later
calculating, we write out the effective pomeron propagator which is 
also radiated from nucleon [3]:
\begin{eqnarray}
\label{22}
D_{I\!\!P}(t) = -3 \beta_{I\!\!P}^2 F_N(t) (\frac{s}{2 m_N^2})^
{\frac{1}{2}(
\alpha_{I\!\!P}(t) -1)}(\frac{1 + e^{-i\pi\alpha_{I\!\!P}(t)}}{sin\pi \alpha_{I\!\!P
}(t)}), \end{eqnarray}
where $\beta_{I\!\!P}\equiv\beta_{I\!\!P}(t)|_{t=0}\approx 1.8 
GeV^{-1}$ is the coupling 
constant between $I\!\!P$ and quarks, 
$\alpha_{I\!\!P}(t)\approx 1.086+ 0.25 t$ is $I\!\!P$-trajectrory expansion in
 small $|t|$,
factor 3 in Eq.(20) corresponds to the 
three quarks in the nucleon.
\begin{center}
\section{$\gamma^* + {I\!\!R_\rho}$ hard process}
\end{center}
\par  
  Both ZEUS and HI Collabs. at the end of 93' in e-p DIS process, observed 
in e-p DIS the large rapidity gap (LRG) phenomena on HERA. Such phenomena were caused 
by those events in which the protons were diffractively scattered 
(or dissociated)
and a pomeron was radiated. Since $I\!\!P$ is a colourless
entity, which is quite different from quark or gluons, colour string would not
attach between X 
(hard subprocess products) and $ X_p$ (the remment of proton),
thus LRG is manifested in the rapidity distribution of final hadrons\footnote{
Of course, since $I\!\!R$ are also colourless entities, LRG would also be 
manifested in the rapidity distribution of final hadrons as that in 
$I\!\!P$ case.}. 
 This is a
further evidence which supports the inference that pomeron participates in hard 
subprocesss and itself has partonic structure.
   
   It seems naturally to ask oneself whether the reggeons, like that case of 
pomeron, could appear in hard subprocess and themself would have partonic 
structure. As a primary test on such problems, we choose the charge-exchanged
photoproduction quark-pair process
 $ \gamma ^{*}+p\rightarrow n+q_u+\bar{q}_d(\rightarrow n+X)$,
as sketched in Fig. 3, which is the simplest hard process induced by 
$I\!\!R _{\rho }$ and we have got the $\rho $-reggeon effective propagator
$D_{I\!\!R_{\rho }}(t)$ and the $I\!\!R_{\rho }~q\overline{q}$ coupling constant
  $\beta _{\rho }$ in the last section. It should be emphasized here that we
  do not in any case want      
to discuss or evaluate the whole story about contributions from 
$I\!\!R _{\rho }$ in DIS process, but only want to estimate the relative
 percents for
contributions from $I\!\!R _{\rho }$ to that from $I\!\!P$, and want to know
whether it can be observed experimentally. As for a comparison, we also 
evaluated the similar contributions both from $I\!\!P$  and partons, 
 as  shown in Fig. 4 and Fig. 5.

    In the follawing we shall write down the relevant formalism sketchily. 
Under the equivalent photo approximation, the cross section for absorption of a
virtual transverse photon, 
$\sigma _{\gamma ^{*}N}^{T}(s,Q^2)$ is related with the
 nucleon structure function $F_2(x,Q^2)$ as
\begin{equation}
F_2(x,Q^2)=\frac{Q^2}{4\pi ^2\alpha }\sigma _{\gamma ^*p}^{T}(s,Q^2) .
\end{equation}
the cross section $\sigma _{\gamma ^*p}^{T}(s,Q^2)$ for Fig. 4 is
\begin{equation}
\sigma _{\gamma^*p}^{T}(s,Q^2)=\frac{1}{4q_{s12}\sqrt{s}}\int\overline{\sum}
|{\cal M}|^2(2 \pi )^4 \delta^4 (p + q - p'-k-k')
\frac{d^3p'}{(2\pi )^32E_{p'}}\frac{d^3k}{(2\pi )^32E_k}
\frac{d^3k'}{(2\pi )^32E_{k'}},
\end{equation}         
 where $s\equiv(p+q)^2$, $q_{s12}\simeq\frac{1}{2\sqrt{s}}(s+Q^2)$.
  After some operations the invariant phase space elements can be transformed to       
\begin{equation}
\frac{d^3p'}{2E'}\simeq\frac{2\pi \sqrt{s}}{s+Q^2}dtd\omega,
\end{equation}
where $t=(p-p')^2$, $\omega =q_0$. Thus we can rewrite Eq.(18) as
\begin{equation}
\sigma _{\gamma ^*p}^{T}(s,Q^2)=\frac{\sqrt{s}}{8(2\pi )^4(s+Q^2)^2}
\int_{-1}^{0}dt\int_{
\frac{2m_{\pi}+Q^2}{2 \sqrt{s}}
}^{\frac{s+Q^2}{10\sqrt{s}}} 
 d\omega \int \frac{d^3k}{E_k}\int \frac{d^3k'}{E_{k'}}\overline{\sum }
\mid {\cal M}\mid^2\delta ^4(q+q'-k-k'),
\end{equation}
In order for comparing conveniently, we have taken the same integal bounds of t
and $\omega $ as in the $I\!\!P$ case, i.e. $-1\leq t\leq 0$, 
$
\frac{2m_{\pi}+Q^2}{2 \sqrt{s}}
\leq \omega \leq \frac{s+Q^2}{10\sqrt{s}}$. According to Eq.(19) 
 and Fig. 4, matrix element ${\cal M}$ is expressed as
\begin{eqnarray}
{\cal M }& = & 2\beta_{\rho }^{2}F_N(t)
(\frac{s}{2m_{N}^{2}})^{\frac{1}{2}
(\alpha_{\rho }(t)-1)}(\frac{1-e^{-i\pi \alpha_{\rho }(t)}}
{sin\pi \alpha _{\rho }(t)})\nonumber \\
&&[Q_u \overline{u_u}(k)\gamma _{\mu }\frac{i(\not\!\!q'-\not\!\!k')}
{(q'-k')^2} \gamma _{\rho }v_d(k')
  +   Q_d\overline{u_u}(k)\gamma _{\rho }\frac{i(\not\!\!q-\not\!\!k')}
{(q-k')^2}\gamma _{\mu }v_d(k')]\overline{u_d}(p')\gamma _{\rho }u_u(p),
\end{eqnarray}
where $Q_u=\frac{2}{3}e$, $Q_d=\frac{-1}{3}e$ for u, d quark charge 
 respectively.
Here we have assumed that in field formalism, the coupling type between 
$\rho $-reggeon and quarks is a vector (i.e. $\gamma _{\rho }$ matrix in 
Eq.(25)), as the same as that in the Pomeron case.\footnote{According to 
Chew-Frautschi plot ,one could alternatively choose the scalar type (I unit 
matrix instead of $ \gamma _{\rho }$ in Eq.(25)), but would lead to a 
result which is 2-3 order of magnitudes smaller than the vetor one.}.

  The operations about $\overline{\sum }\mid {\cal M}\mid ^2$ are standard, 
and it is completed in computer by using the Reduce Programm.
   
   We have also considered the pomeron contributions which come from Fig. 4 to
$\sigma _{\gamma ^{*}p}^{T}(s,Q^2)$ $(or F_2(x,Q^2))$, here the formalism is 
 almost the same as that of $\rho $-reggeon, except the effective propagator 
$D_{I\!\!R_{\rho }}$(t) in Eq.(19) is replaced by $D_{I\!\!P }$(t) in Eq.(20) 
 and change some trivial factors such as coupling constant $\beta$,  quark charge etc.  As we have 
 mentioned, we need also evaluate contributions in $O(\alpha _{s})$ from 
ordinary photo-partons interaction, its diagrams are described in Fig. 5. This 
 could be easily done since everything is known and ready. Ref. \cite{7a}
 has given out  the next 
 leading order related 
 formula
\begin{equation}
\frac{1}{x}F_2(x,Q^2)=\sum e_{q}^{2}\{q(x,Q^2)+\overline{q}(x,Q^2)+
\frac{\alpha _s(Q^2)}{2\pi }
[C_{q,2}\otimes (q+\overline{q})+C_{g,2}\otimes g]\},
\end{equation}
where $C_{q,2}(z)$ and $C_{g,2}(z)$ 
are structure functions of quark and gloun at
$O(\alpha_s)$ respectively,
\begin{equation}
C_{q,2}(z)=\frac{4}{3}\{\frac{1+z^2}{1-z^2}[ln(\frac{1-z}{z})-\frac{3}{4}]+
\frac{1}{4}(9+5z)\}_{+},
\end{equation}
\begin{equation}  
C_{g,2}(z)=\frac{1}{2}\{[z^2+(1-z)^2]ln\frac{1-z}{z}-1+8z(1-z)\}.
\end{equation}
$C_{q,2}\otimes (q+\overline{q})$ and $C_{g,2}\otimes g$ are short
hand notations of following convolutions respectively,
\begin{eqnarray}
C_{q,2}\otimes (q+\overline{q})\equiv\int_{x}^{1}\frac{dy}{y}C_{q,2}
(\frac{x}{y})[q(y,Q^2)+\overline{q}(y,Q^2)],\nonumber \\
C_{g,2}\otimes g\equiv\int_{x}^{1}\frac{dy}{y}C_{g,2}
(\frac{x}{y})g(y,Q^2),
\end{eqnarray}
where the $\overline{MS}$ renormalization scheme has been used. As for the
parton distribution functions, $q(x,Q^2)$, $\overline{q}(x,Q^2)$ and
$g(x,Q^2)$ needed in Eq.(26), we take them from [8].

%\vfill\eject\pagestyle{plain}\setcounter{page}{1}
\begin{center}
\section{Results and Conclusions}
\end{center}

Since in this paper we want to study the possible contributions of Reggeon in
hard process of strong interaction and such work had never been done by others.
We must start out to get the effective propagator of a reggeon, $D_{I\!\!R}(t)$
and the coupling  constant $\beta$ between reggeon and partons.Thus we choose
an ideal process, $\pi^-+p\rightarrow\pi^0+n$, in which only the $\rho$-reggeon
could be exchanged. From the related phenomenological analysis and under some
ansatzs, we have got the $\rho$-reggeon effective propagator $D_{I\!\!R_{\rho}}$(t) 
and coupling constant between $\rho$-reggeon with light quark, 
$\beta_{\rho}$(t). Put $D_{I\!\!R_{\rho}}$(t) and $\beta_{\rho}$ in comparing with the 
corresponding $D_{I\!\!P}$(t) and $\beta_{I\!\!P}$ of pomeron, we see they 
are reasonable.

For the $\rho$-reggeon contribution in hard process aspects, as a simple example
we have discussed the 
 charge-exchanged photonproduction quark-pair process,
$\gamma^*+p\rightarrow n+q_u+\overline{q}_d(\rightarrow n+X)$, which may be tested 
at HERA. The kinematical regions covered by our calculation for above process
and its controbutions to structure function $F_2(x,Q^2)$ in e-p DIS process
are shown in Fig. 6. As for comparison, contributions from the process as
described by Eq.(3) and (4) for pomeron and ordinary partons are also shown
in it. From Fig. 6 one can see that:
\begin{enumerate}
\item The ratio of $F_2(x,Q^2)$ from $I\!\!R_{\rho}$
to that from $I\!\!P$ is larger  than  
one percent from small $x$ to large $x$ range when $Q^2=10~ GeV^2$. 
 This is what we expected.
Since from Chew-Frautschi plot in Fig. 1 and Eqs.(19), (20) and (24), we see
factors $(\frac{s}{2m_{N}^{2}})^{\alpha_{I\!\!R_{\rho}}(t)-1}$ and
$(\frac{s}{2m_{N}^{2}})^{\alpha_{I\!\!P}(t)-1}$ which are critical ingredients
for cross sections of $I\!\!R_{\rho}$ and $I\!\!P$, the ratio between
them is $e^{-0.586ln(\frac{s}{2})}$, which is ${\cal O}(10^{-2})$,
in our main kinematical regions.
\item The {\it standard} parton  contributions (Fig. 5) to $F_2(x,Q^2)$
are  $1 \sim 2$  orders 
  of magnitude larger than that of $I\!\!P$,
but we can see its major  contributions come from leading
order (i.e. from the modified naive parton model with $q(x,Q^2)$ instead of
q(x)). Results show that all next leading order  corrections
to $F_2 (x,Q^2)$ are about, on the average, the same order as 
that from $I\!\!P$.
\end{enumerate}

Since the $F_2(x,Q^2)$ of $I\!\! R _\rho$ is $1\sim 2$ orders smaller than 
that of $I\!\! P$, we expect that the $I\!\! R_\rho$ signals may be
tested
at HERA  in hard process
such as Fig. 3. But as prerequisite, there must be a sensitive neutron 
detector to detect the final forward neutrons. Since the critical step to
distinguish events coming from $I\!\!R_{\rho}$-exchanged or from $I\!\!P$-exchanged
is to observe whether the detected final forward hardron is a neutron or
a proton. Recently, ZEUS Collab. in DESY have installed on HERA a forward proton
spectrometer(FPS) and a forward neutron calorimeter (FNCAL) [9] which perhaps
would be feasible for above test.
\begin{centerline}{\bf Acknowledgement}\end{centerline}
We are grateful to Feng Yuan and Jiasheng Xu for their kindly help.

\end{document}